# OBSERVATION OF THE $\Xi_c^+$ CHARMED BARYON DECAYS TO $\Sigma^+K^-\pi^+$, $\Sigma^+\bar{K}^{*0}$, and $\Lambda K^-\pi^+\pi^+$

T. Bergfeld,[1] B.I. Eisenstein,[1] J. Ernst,[1] G.E. Gladding,[1] G.D. Gollin,[1] M. Palmer,[1] M. Selen,[1] J.J. Thaler,[1] K.W. Edwards,[2] K.W. McLean,[2] M. Ogg,[2] A. Bellerive,[3] D.I. Britton,[3] E.R.F. Hyatt,[3] R. Janicek,[3] D.B. MacFarlane,[3] P.M. Patel,[3] B. Spaan,[3] A.J. Sadoff,[4] R. Ammar,[5] P. Baringer,[5] A. Bean,[5] D. Besson,[5] D. Coppage,[5] N. Copty,[5] R. Davis,[5] N. Hancock,[5] S. Kotov,[5] I. Kravchenko,[5] N. Kwak,[5] Y. Kubota,[6] M. Lattery,[6] M. Momayezi,[6] J.K. Nelson,[6] S. Patton,[6] R. Poling,[6] V. Savinov,[6] S. Schrenk,[6] R. Wang,[6] M.S. Alam,[7] I.J. Kim,[7] Z. Ling,[7] A.H. Mahmood,[7] J.J. O'Neill,[7] H. Severini,[7] C.R. Sun,[7] F. Wappler,[7] G. Crawford,[8] J.E. Duboscq,[8] R. Fulton,[8] D. Fujino,[8] K.K. Gan,[8] K. Honscheid,[8] H. Kagan,[8] R. Kass,[8] J. Lee,[8] M. Sung,[8] C. White,[8] A. Wolf,[8] M.M. Zoeller,[8] X. Fu,[9] B. Nemati,[9] W.R. Ross,[9] P. Skubic,[9] M. Wood,[9] M. Bishai,[10] J. Fast,[10] E. Gerndt,[10] J.W. Hinson,[10] T. Miao,[10] D.H. Miller,[10] M. Modesitt,[10] E.I. Shibata,[10] I.P.J. Shipsey,[10] P.N. Wang,[10] L. Gibbons,[11] S.D. Johnson,[11] Y. Kwon,[11] S. Roberts,[11] E.H. Thorndike,[11] T.E. Coan,[12] J. Dominick,[12] V. Fadeyev,[12] I. Korolkov,[12] M. Lambrecht,[12] S. Sanghera,[12] V. Shelkov,[12] T. Skwarnicki,[12] R. Stroynowski,[12] I. Volobouev,[12] G. Wei,[12] M. Artuso,[13] M. Gao,[13] M. Goldberg,[13] D. He,[13] N. Horwitz,[13] S. Kopp,[13] G.C. Moneti,[13] R. Mountain,[13] F. Muheim,[13] Y. Mukhin,[13] S. Playfer,[13] S. Stone,[13] X. Xing,[13] J. Bartelt,[14] S.E. Csorna,[14] V. Jain,[14] S. Marka,[14] D. Gibaut,[15] K. Kinoshita,[15] P. Pomianowski,[15] B. Barish,[16] M. Chadha,[16] S. Chan,[16] D.F. Cowen,[16] G. Eigen,[16] J.S. Miller,[16] C. O'Grady,[16] J. Urheim,[16] A.J. Weinstein,[16] F. Würthwein,[16] D.M. Asner,[17] M. Athanas,[17] D.W. Bliss,[17] W.S. Brower,[17] G. Masek,[17] H.P. Paar,[17] J. Gronberg,[18] C.M. Korte,[18] R. Kutschke,[18] S. Menary,[18] R.J. Morrison,[18] S. Nakanishi,[18] H.N. Nelson,[18] T.K. Nelson,[18] C. Qiao,[18] J.D. Richman,[18] D. Roberts,[18] A. Ryd,[18] H. Tajima,[18] M.S. Witherell,[18] R. Balest,[19] K. Cho,[19] W.T. Ford,[19] M. Lohner,[19] H. Park,[19] P. Rankin,[19] J.G. Smith,[19] J.P. Alexander,[20] C. Bebek,[20] B.E. Berger,[20] K. Berkelman,[20] K. Bloom,[20] T.E. Browder,[20*] D.G. Cassel,[20] H.A. Cho,[20] D.M. Coffman,[20] D.S. Crowcroft,[20] M. Dickson,[20] P.S. Drell,[20] D.J. Dumas,[20] R. Ehrlich,[20] R. Elia,[20] P. Gaidarev,[20] M. Garcia-Sciveres,[20] B. Gittelman,[20] S.W. Gray,[20] D.L. Hartill,[20] B.K. Heltsley,[20] S. Henderson,[20] C.D. Jones,[20] S.L. Jones,[20] J. Kandaswamy,[20] N. Katayama,[20] P.C. Kim,[20] D.L. Kreinick,[20] T. Lee,[20] Y. Liu,[20] G.S. Ludwig,[20] J. Masui,[20] J. Mevissen,[20] N.B. Mistry,[20] C.R. Ng,[20] E. Nordberg,[20] J.R. Patterson,[20] D. Peterson,[20] D. Riley,[20] A. Soffer,[20] P. Avery,[21] A. Freyberger,[21] K. Lingel,[21] C. Prescott,[21] J. Rodriguez,[21] S. Yang,[21] J. Yelton,[21] G. Brandenburg,[22] D. Cinabro,[22] T. Liu,[22] M. Saulnier,[22] R. Wilson,[22] and H. Yamamoto[22]

(CLEO Collaboration)
[1] *University of Illinois, Champaign-Urbana, Illinois, 61801*
[2] *Carleton University, Ottawa, Ontario K1S 5B6 and the Institute of Particle Physics, Canada*
[3] *McGill University, Montréal, Québec H3A 2T8 and the Institute of Particle Physics, Canada*
[4] *Ithaca College, Ithaca, New York 14850*
[5] *University of Kansas, Lawrence, Kansas 66045*
[6] *University of Minnesota, Minneapolis, Minnesota 55455*
[7] *State University of New York at Albany, Albany, New York 12222*
[8] *Ohio State University, Columbus, Ohio, 43210*
[9] *University of Oklahoma, Norman, Oklahoma 73019*
[10] *Purdue University, West Lafayette, Indiana 47907*
[11] *University of Rochester, Rochester, New York 14627*
[12] *Southern Methodist University, Dallas, Texas 75275*
[13] *Syracuse University, Syracuse, New York 13244*
[14] *Vanderbilt University, Nashville, Tennessee 37235*
[15] *Virginia Polytechnic Institute and State University, Blacksburg, Virginia, 24061*
[16] *California Institute of Technology, Pasadena, California 91125*
[17] *University of California, San Diego, La Jolla, California 92093*
[18] *University of California, Santa Barbara, California 93106*
[19] *University of Colorado, Boulder, Colorado 80309-0390*
[20] *Cornell University, Ithaca, New York 14853*
[21] *University of Florida, Gainesville, Florida 32611*
[22] *Harvard University, Cambridge, Massachusetts 02138*

## Abstract

We have observed two new decay modes of the charmed baryon $\Xi_c^+$ into $\Sigma^+K^-\pi^+$ and $\Sigma^+\bar{K}^{*0}$ using data collected with the CLEO II detector. We also present the first measurement of the branching fraction for the previously observed decay mode $\Xi_c^+ \to \Lambda K^-\pi^+\pi^+$. The branching fractions for these three modes relative to $\Xi_c^+ \to \Xi^-\pi^+\pi^+$ are measured to be $1.18 \pm 0.26 \pm 0.17$, $0.92 \pm 0.27 \pm 0.14$, and $0.58 \pm 0.16 \pm 0.07$, respectively.

---

*Permanent address: University of Hawaii at Manoa

The $\Xi_c^+$ charmed baryon lifetime is 2.5 to 3.5 times longer than the $\Xi_c^0$ lifetime [1,2]. It is widely believed that destructive interference between the external and internal W-emission diagrams in $\Xi_c^+$ decays and the presence of W-exchange decay channels in $\Xi_c^0$ decays contribute to the difference in lifetimes [3]. Measuring the branching fractions of $\Xi_c$ decay modes can help verify these assertions and constrain the mechanisms in $\Xi_c$ charmed baryon decays.

Although numerous decay modes of the $\Lambda_c^+$ charmed baryon have been observed, only a few $\Xi_c^+$ decay modes have been reported. The $\Xi_c^+ \to \Xi^-\pi^+\pi^+$ and $\Lambda K^-\pi^+\pi^+$ decay modes [4] have been previously observed by a number of groups [1,5]. CLEO has also observed the $\Xi_c^+$ decaying into modes with a neutral $\Xi^0$ hyperon, namely, $\Xi_c^+ \to \Xi^0\pi^+$, $\Xi^0\pi^+\pi^0$, and $\Xi^0\pi^+\pi^-\pi^+$ [6]. The simplest external W spectator diagram produces a $\Xi^0$ in the final state. Other hyperons ($\Lambda$, $\Sigma^+$, and $\Xi^-$) can be produced either by internal W-emission diagrams or by combining valence quarks with quarks popped from the vacuum.

We report in this Letter the observation of two new decay modes of the $\Xi_c^+$ into $\Sigma^+K^-\pi^+$ and $\Sigma^+\bar{K}^{*0}$. The simplest diagram yielding $\Sigma^+\bar{K}^{*0}$ is an internal W-emission diagram. The $\Sigma^+K^-\pi^+$ mode has been previously seen by the CERN fixed target experiment ACCMOR [7], but is based on three events in the mass range $2445 - 2475$ MeV/c$^2$. We also present a first measurement of the $\Xi_c^+ \to \Lambda K^-\pi^+\pi^+$ branching fraction.

The data were collected with the CLEO II detector at the Cornell $e^+e^-$ storage ring CESR, which operated on, and just below, the $\Upsilon(4S)$ resonance. The CLEO II detector [8] is a large solenoidal detector with 67 tracking layers and a CsI electromagnetic calorimeter that provides efficient $\pi^0$ reconstruction. We have used a total integrated luminosity of 3.6 fb$^{-1}$, which corresponds to roughly 4 million $c\bar{c}$ events.

Samples of $\Lambda \to p\pi^-$, $D^0 \to K^-\pi^+$ from $D^{*+}$'s, and $K_s^0 \to \pi^+\pi^-$ in the data are used to measure the particle identification efficiencies for protons, kaons, and pions, respectively. Charged proton and pion candidates are required to have specific ionization loss ($dE/dx$) information and, when available, time-of-flight information consistent with the value expected for the assumed particle type. The proton in our $\Xi_c^+$ decay modes always comes from a hyperon ($\Lambda$, $\Sigma^+$, or $\Xi^-$) and backgrounds from other hadrons are relatively low. Charged kaons, however, are selected with stronger particle identification cuts to minimize the reflection peaks in the $\Xi_c^+$ invariant mass distributions from the $\Lambda_c^+ \to \Lambda\pi^+\pi^-\pi^+$ and $\Sigma^+\pi^-\pi^+$ decay modes, where one of the pions is misidentified as a kaon. These reflections will produce broad distributions peaked near the $\Xi_c^+$ mass. We require the probability that the candidate is a kaon to be at least 50% of the sum of the probabilities for the proton, kaon, and pion hypotheses.

The $\Lambda$ candidates are selected in their decay $\Lambda \to p\pi^-$ by reconstructing a secondary decay vertex from the intersection of two oppositely charged tracks in the plane perpendicular to the beam axis (the $r-\phi$ plane). Similarly, $\Xi^-$ hyperons are selected through the decay mode $\Xi^- \to \Lambda\pi^-$ by reconstructing a secondary decay vertex from the intersection of a $\Lambda$ candidate and a $\pi^-$ particle in the $r-\phi$ plane. The $\Sigma^+$ hyperon candidates are selected from $p\pi^0$ combinations that are consistent with originating from a decay vertex displaced from the primary interaction point [9]. The invariant masses of the $\Lambda$, $\Xi^-$, and $\Sigma^+$ candidates must lie within 5, 6, and 15 MeV/c$^2$ of their nominal values ($\sim 3\sigma$), respectively.

Charmed baryons from $e^+e^- \to c\bar{c}$ interactions are produced with a hard momentum spectrum, so the combinatoric background can be reduced by requiring either $x_p > 0.5$ or $x_p > 0.6$, depending on the decay mode, where $x_p = P_{\Xi_c^+}/\sqrt{E_{beam}^2 - M_{\Xi_c^+}^2}$ is the scaled momentum of the $\Xi_c^+$. The $x_p$ cut also eliminates $\Xi_c^+$ baryons that arise from $B$ meson decays. In addition, we require that the decay products of the $\Xi_c^+$ lie within 90 degrees of the candidate $\Xi_c^+$ momentum vector.

The invariant mass distribution for $\Xi_c^+ \to \Sigma^+ K^- \pi^+$ candidates with $x_p > 0.5$ is shown in Figure 1. We observe a clear $\Xi_c^+$ signal at $\sim 2470$ MeV/c$^2$. However, the distribution also contains a reflection peak from $\Lambda_c^+ \to \Sigma^+ \pi^- \pi^+$ events where a pion is misidentified as a kaon. The shape of the reflection peak is determined from a Monte Carlo phase space decay of $\Lambda_c^+ \to \Sigma^+ \pi^- \pi^+$. The area is determined from data by constructing the invariant mass distribution for the $\Sigma^+ \pi^- \pi^+$ hypothesis from all $\Sigma^+ K^- \pi^+$ combinations in Figure 1. A $\Lambda_c^+ \to \Sigma^+ \pi^- \pi^+$ signal of $142 \pm 25$ events is observed, which is taken to be the area of the reflection peak.

We parametrize the $\Sigma^+ K^- \pi^+$ mass distribution by a Gaussian signal, the $\Lambda_c^+ \to \Sigma^+ \pi^- \pi^+$ reflection peak with normalization fixed to 142 events, and a 3rd order Chebyshev polynomial background. The width of the Gaussian is determined from Monte Carlo studies to be $\sigma = 8.6$ MeV/c$^2$. We observe $119 \pm 23$ events (statistical error only) in the $\Xi_c^+ \to \Sigma^+ K^- \pi^+$ decay mode at a mass of $2469.9 \pm 2.0$ MeV/c$^2$ (statistical error only), consistent with the nominal $\Xi_c^+$ mass. Excluding the $\Lambda_c^+$ reflection in the fit artificially raises the $\Xi_c^+$ yield.

A search was made for the two-body decay $\Xi_c^+ \to \Sigma^+ \bar{K}^{*0}$ by examining the resonant substructure of the $\Sigma^+ K^- \pi^+$ mode. We divide the data shown in Figure 1 into seven regions of $M_{K^-\pi^+}$ (of unequal size) from 0.6 to 1.3 GeV/c$^2$ and obtain the $\Xi_c^+$ yield for each region. Contributions from the $\Lambda_c^+$ reflection are again measured and incorporated in the fit. Figure 2 shows a clear $\bar{K}^{*0}$ signal in the $K^- \pi^+$ invariant mass distribution. We fit the data (points) to the sum of two curves representing the resonant and non-resonant (n.r.) $\Sigma^+ K^- \pi^+$ contributions (histogram), where the shapes but not the normalizations are determined from Monte Carlo. The fit yields $61 \pm 17$ events from $\Xi_c^+ \to \Sigma^+ \bar{K}^{*0}$ and $55 \pm 22$ events from n.r. $\Xi_c^+ \to \Sigma^+ K^- \pi^+$ decays. The efficiency corrected yields are $619 \pm 172$ events for the $\Xi_c^+ \to \Sigma^+ \bar{K}^{*0}$, $\bar{K}^{*0} \to K^- \pi^+$ channel and $525 \pm 208$ events for the n.r. $\Xi_c^+ \to \Sigma^+ K^- \pi^+$ mode.

As a check, we also measure the $\Xi_c^+ \to \Sigma^+ \bar{K}^{*0}$ contribution by examining the $\Sigma^+ K^- \pi^+$ invariant mass distribution after requiring the $K^- \pi^+$ mass to be within 50 MeV/c$^2$ of the $\bar{K}^{*0}$ mass (see Figure 3). About half the $\Xi_c^+$ signal remains and the backgrounds are greatly reduced. A fit to the $\Sigma^+ K^- \pi^+$ mass distribution yields $59 \pm 12$ events. We have included the $\Sigma^+ \pi^- \pi^+$ reflection peak with a fixed area of 35 events, as determined in the data. Since the $\bar{K}^{*0}$ mass cut is wide, a non-negligible amount of non-resonant $\Xi_c^+ \to \Sigma^+ K^- \pi^+$ decays can survive the cut, although the reconstruction efficiency is a factor of four smaller than that of resonant $\Sigma^+ \bar{K}^{*0}$. We obtain a nearly pure sample of n.r. $\Sigma^+ K^- \pi^+$ by requiring that $|M_{K\pi} - 892| > 100$ MeV/c$^2$. In this case, we measure $45 \pm 17$ events in the data. After unfolding the resonant and n.r. contributions from these two measurements, we obtain an efficiency corrected yield of $669 \pm 179$ events for the decay chain $\Xi_c^+ \to \Sigma^+ \bar{K}^{*0}$, $\bar{K}^{*0} \to K^- \pi^+$ and $550 \pm 252$ events for the non-resonant mode $\Xi_c^+ \to \Sigma^+ K^- \pi^+$. This is consistent with the primary result and the difference is taken as a systematic error.

The invariant mass distribution for $\Xi_c^+ \to \Lambda K^- \pi^+ \pi^+$ candidates with $x_p > 0.6$ is shown in Figure 4. The more stringent $x_p$ requirement is needed to reduce the combinatoric back-ground for this higher multiplicity final state. We parametrize the mass distribution as the sum of a Gaussian signal, a contribution from the $\Lambda_c^+ \to \Lambda\pi^+\pi^-\pi^+$ reflection peak, and a 3rd order Chebyshev polynomial background. The width of the Gaussian is taken from Monte Carlo studies to be $\sigma = 7.0$ MeV/c$^2$. The reflection peak is fixed to the $\Lambda_c^+ \to \Lambda\pi^+\pi^-\pi^+$ yield of $116 \pm 22$ events, determined from a fit to the invariant mass distribution for the $\Lambda\pi^+\pi^-\pi^+$ hypothesis for all $\Lambda K^-\pi^+\pi^+$ combinations in Figure 4.

We observe $61 \pm 15$ events for the mode $\Xi_c^+ \to \Lambda K^-\pi^+\pi^+$ at a mass of $2467.5 \pm 2.0$ MeV/c$^2$. We have also looked for possible $\bar{K}^{*0}$ and $\Sigma^{*+}(1385)$ resonant contributions in $\Xi_c^+ \to \Lambda K^-\pi^+\pi^+$ decays. None is observed, and upper limits at the 90% confidence level of $\mathcal{B}(\Lambda\bar{K}^{*0}\pi^+)/\mathcal{B}(\Lambda K^-\pi^+\pi^+) < 0.5$ and $\mathcal{B}(\Sigma^{*+}K^-\pi^+)/\mathcal{B}(\Lambda K^-\pi^+\pi^+) < 0.7$ are obtained.

Since the $\Xi_c^+$ total cross section is not precisely known, we convert our observations into branching fractions relative to the well measured decay mode $\Xi_c^+ \to \Xi^-\pi^+\pi^+$. For each mode we apply the same $\Xi_c^+$ momentum cut to $\Xi^-\pi^+\pi^+$ (either 0.5 or 0.6) to reduce the systematic dependence on the $x_p$ cut. The invariant mass distribution for $\Xi_c^+ \to \Xi^-\pi^+\pi^+$ candidates with $x_p > 0.5$ is shown in Figure 5. We parametrize the mass distribution by a Gaussian signal and a quadratic polynomial background. The width of the Gaussian is taken from Monte Carlo studies to be $\sigma = 10.1$ MeV/c$^2$. We observe $131 \pm 14$ events at a mass of $2466.6 \pm 1.3$ MeV/c$^2$. For $x_p > 0.6$ we observe $100 \pm 11$ $\Xi_c^+$ events. No resonant substructure is observed, and we place an upper limit for the two-body decay $\Xi_c^+ \to \Xi_c^{*0}(1530)\pi^+$ to be $\mathcal{B}(\Xi_c^+ \to \Xi_c^{*0}\pi^+)/\mathcal{B}(\Xi_c^+ \to \Xi^-\pi^+\pi^+) < 0.2$ at the 90% confidence level. Körner and Krämer [10] have predicted that this decay rate should be zero.

The raw yields, efficiencies, and resultant branching fractions for all decay modes are shown in Table I. The main sources of systematic error are due to uncertainties in the efficiencies for $\Lambda$, $\Sigma^+$, and $\Xi^-$ reconstruction ($5-7\%$), particle identification (5%), charged particle tracking (4%), and $\pi^0$ reconstruction (5%). An additional 8% error is assigned to the $\Sigma^+\bar{K}^{*0}$ yield for the difference between the two methods of extracting the $\bar{K}^{*0}$ contribution. Variations in the $\Xi_c^+$ yields due to uncertainties in the signal width are $2-6\%$ and in the area of the reflection peak for $\Lambda_c^+ \to \Lambda\pi^+\pi^-\pi^+$ and $\Sigma^+\pi^-\pi^+$ are $2-5\%$, depending on the channel. CLEO observes that about half of the $\Lambda_c^+ \to \Lambda\pi^+\pi^-\pi^+$ decays proceed via a $\Sigma^{*+}$ or $\Sigma^{*-}$ resonance [11]. Variations in the shape of the $\Lambda_c^+$ reflection peak due to the $\Sigma^{*\pm}$ substructure correspond to a 3% systematic error. The total systematic error for the $\Sigma^+K^-\pi^+$, $\Sigma^+\bar{K}^{*0}$, and $\Lambda K^-\pi^+\pi^+$ decay modes are 14%, 15%, and 12%, respectively.

Körner and Krämer [10], Zenczykowski [12], and Datta [13] have used quark and pole models to make theoretical predictions for the two-body decay $\Xi_c^+ \to \Sigma^+\bar{K}^{*0}$. They calculate $\Gamma(\Xi_c^+ \to \Sigma^+\bar{K}^{*0})$ to be $5.3 \times 10^{10}s^{-1}$, $8.7 \times 10^{10}s^{-1}$, and $3.6 \times 10^{10}s^{-1}$, respectively. We can convert our measured branching fraction into a decay rate by using the world average $\Xi_c^+$ lifetime [1] and CLEO's measurement of $\mathcal{B}(\Xi_c^+ \to \Xi^-\pi^+\pi^+) = f_{\Xi_c}(2.1 \pm 0.8 \pm 0.4)\%$ [2], where the fraction $f_{\Xi_c} \equiv \mathcal{B}(\Xi_c^+ \to \Xi^0\ell\nu)/\mathcal{B}(\Xi_c^+ \to \ell X)$ is expected to lie in the range $0.6-1.0$. We thus obtain $\Gamma(\Xi_c^+ \to \Sigma^+\bar{K}^{*0}) = f_{\Xi_c}(5.5 \pm 2.9) \times 10^{10}s^{-1}$, in good agreement with the theoretical predictions.

In summary, we have observed two new decay modes, $\Xi_c^+ \to \Sigma^+K^-\pi^+$ and $\Xi_c^+ \to \Sigma^+\bar{K}^{*0}$. We also report the first measurement of the relative branching fraction for $\Xi_c^+ \to \Lambda K^-\pi^+\pi^+$. The ratios of branching fractions suggest that the two strange quarks in the final state do not always hadronize to form a $\Xi$ hyperon, but can instead form two separate hadrons via an internal W-emission diagram. Our measurement of $\mathcal{B}(\Lambda K^-\pi^+\pi^+)/\mathcal{B}(\Xi^-\pi^+\pi^+)$ is consid-erably lower than the preliminary estimate by the WA89 experiment of $\approx 4$ [5]. Somewhat surprisingly, neither $\Lambda K^- \pi^+ \pi^+$ or $\Xi^- \pi^+ \pi^+$ have any observable resonant substructure. This behavior is similar to the $\Lambda_c^+$ charmed baryon in which only a quarter of the observed $\Lambda_c^+$ hadronic width (about 40% of the width has been accounted for) goes through two-body decay modes, with the bulk proceeding through multibody states. In contrast, about half of the hadronic width of $D^+$ and $D^0$ mesons proceeds via two-body decays. Clearly, charmed baryon decays continue to be an interesting laboratory for weak decay physics.

TABLE I. Summary of results on new $\Xi_c^+$ decay modes. The $\Sigma^+ K^- \pi^+$ mode includes both resonant and non-resonant contributions. The efficiencies ($\mathcal{E}$) do not include branching fractions to the observed final states. The first error in the relative branching fraction is statistical and the second is systematic.

| Decay Mode | $x_p$ cut | Events | $\mathcal{E}$ (%) | $\mathcal{B}/\mathcal{B}(\Xi_c^+ \to \Xi^- \pi^+ \pi^+)$ |
|---|---|---|---|---|
| $\Sigma^+ K^- \pi^+$ | 0.5 | $119 \pm 23$ | 10.4 | $1.18 \pm 0.26 \pm 0.17$ |
| $\Sigma^+ \bar{K}^{*0}$ | 0.5 | $61 \pm 17$ | 9.8 | $0.92 \pm 0.27 \pm 0.14$ |
| $\Lambda K^- \pi^+ \pi^+$ | 0.6 | $61 \pm 15$ | 11.5 | $0.58 \pm 0.16 \pm 0.07$ |
| $\Xi^- \pi^+ \pi^+$ | 0.5 | $131 \pm 14$ | 10.6 | 1.0 |

We gratefully acknowledge the effort of the CESR staff in providing us with excellent luminosity and running conditions. This work was supported by the National Science Foundation, the U.S. Department of Energy, the Heisenberg Foundation, the Alexander von Humboldt Stiftung, the Natural Sciences and Engineering Research Council of Canada, and the A.P. Sloan Foundation.

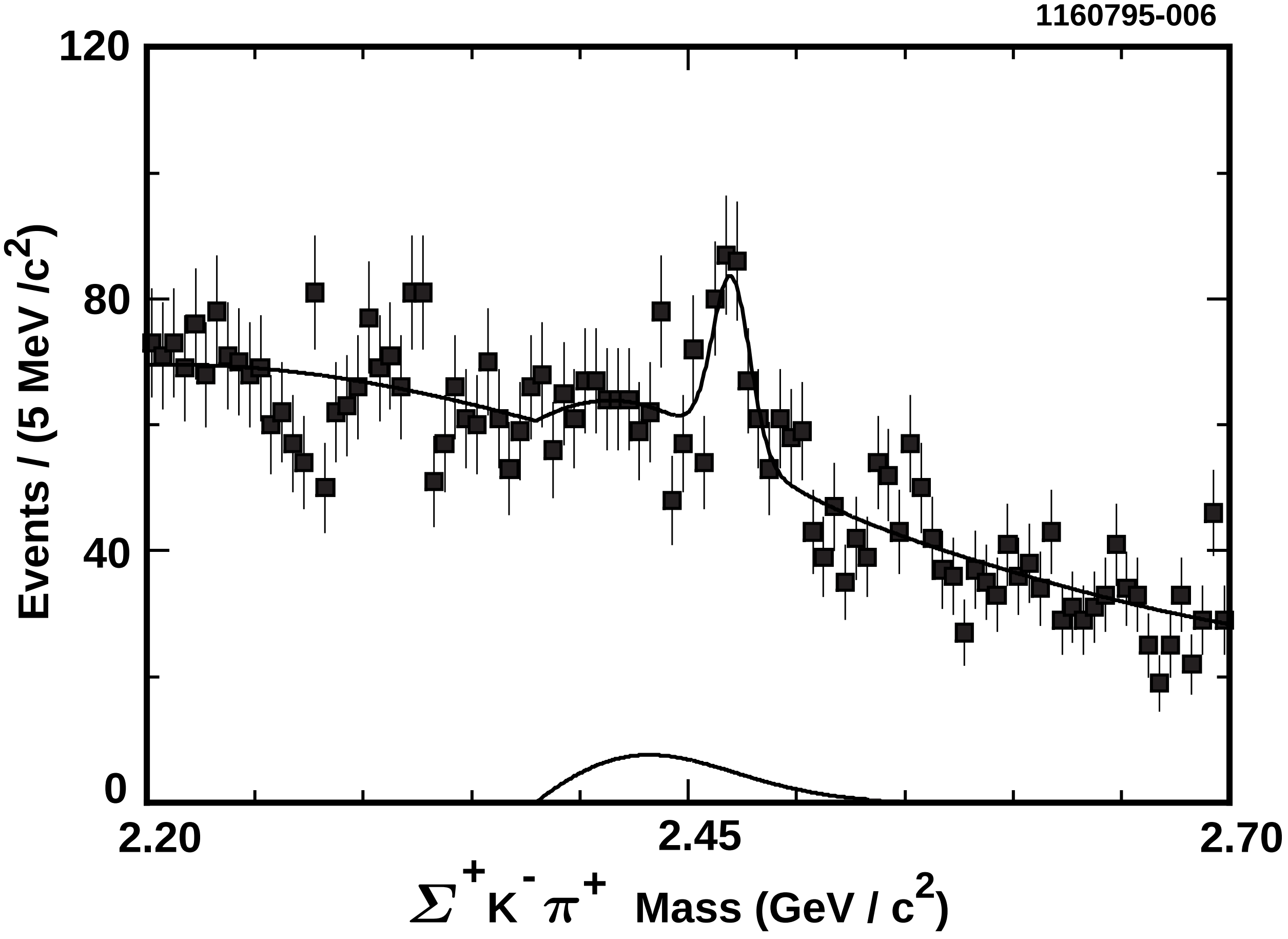

FIG. 1. Invariant mass distribution for $\Xi_c^+ \to \Sigma^+ K^- \pi^+$. The curve at the bottom corresponds to the reflection peak from misidentified $\Lambda_c^+ \to \Sigma^+ \pi^- \pi^+$ events.

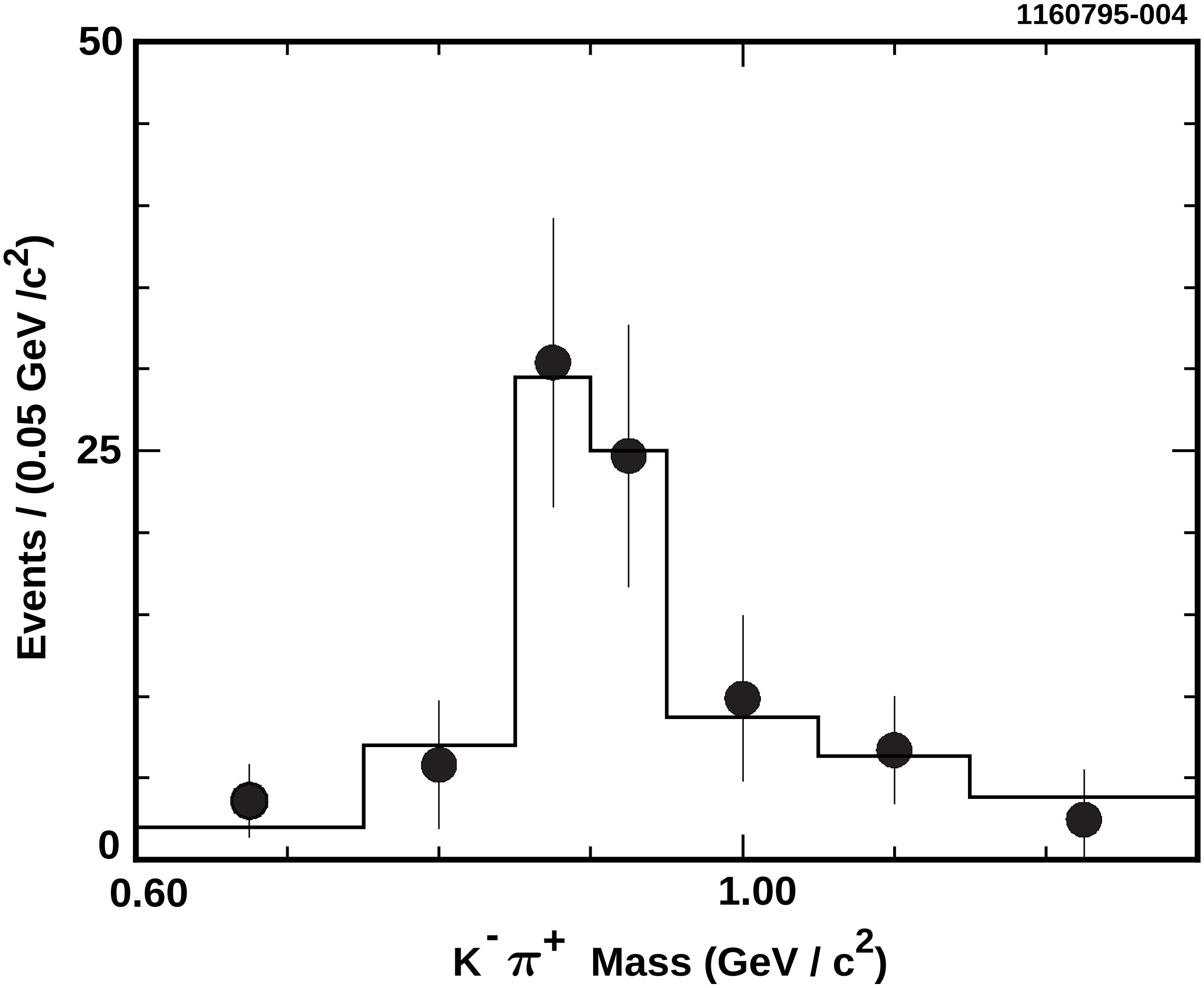

FIG. 2. The $K^-\pi^+$ invariant mass distribution for the $\Xi_c^+ \to \Sigma^+ K^- \pi^+$ decay mode. The points are from data, where the fitted $\Xi_c^+ \to \Sigma^+ K^- \pi^+$ yield is plotted for each bin of $K^-\pi^+$ mass. The histogram is a $\chi^2$ fit to the sum of two curves representing the Monte Carlo shapes for resonant $\Sigma^+ \bar{K}^{*0}$ and non-resonant $\Sigma^+ K^- \pi^+$.

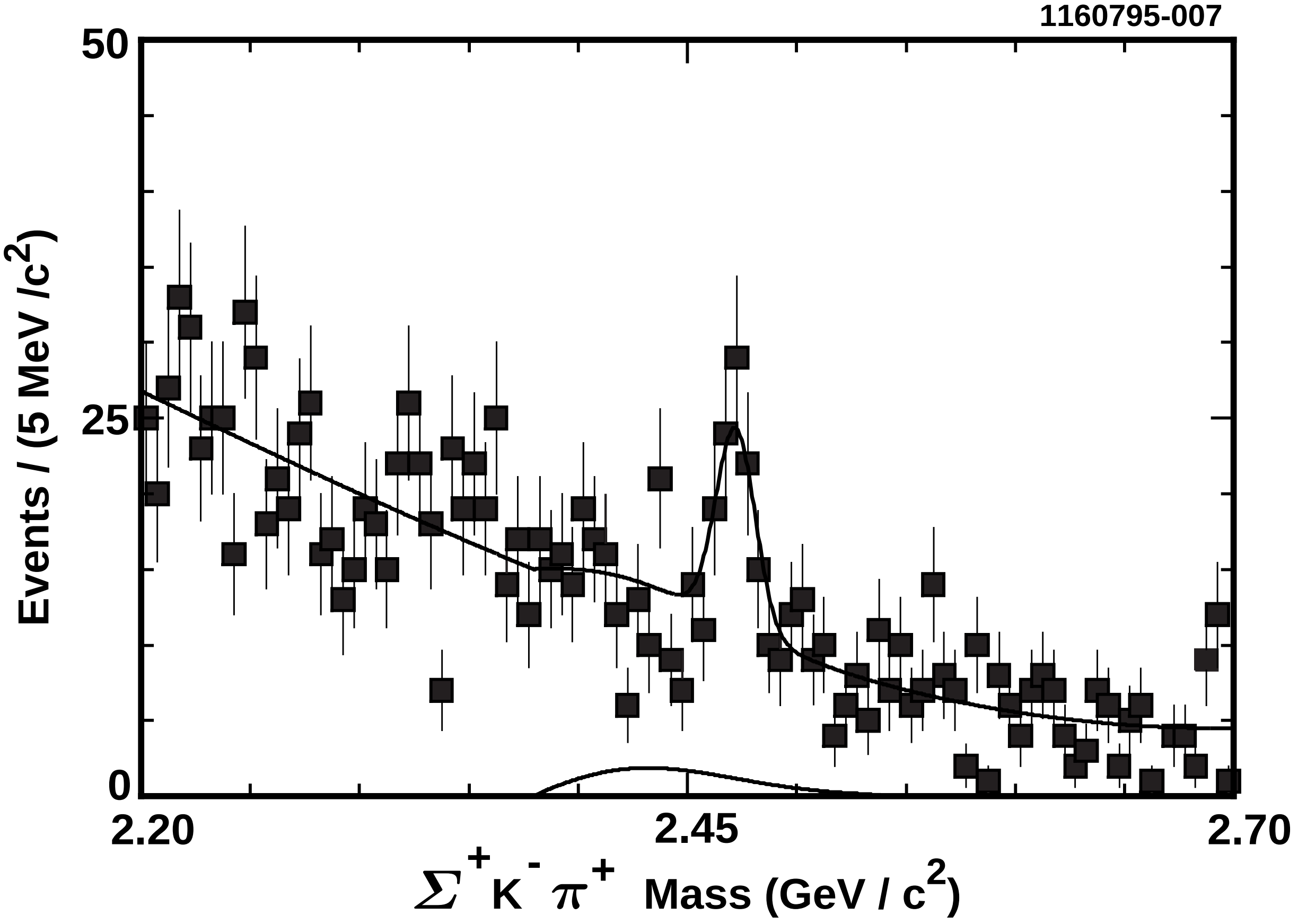

FIG. 3. Invariant mass distribution for $\Xi_c^+ \to \Sigma^+ K^- \pi^+$ with the $\bar{K}^{*0}$ mass cut of $|M_{K\pi} - 892| < 50$ MeV/c$^2$. The curve at the bottom corresponds to the reflection peak from misidentified $\Lambda_c^+ \to \Sigma^+ \pi^- \pi^+$ events.

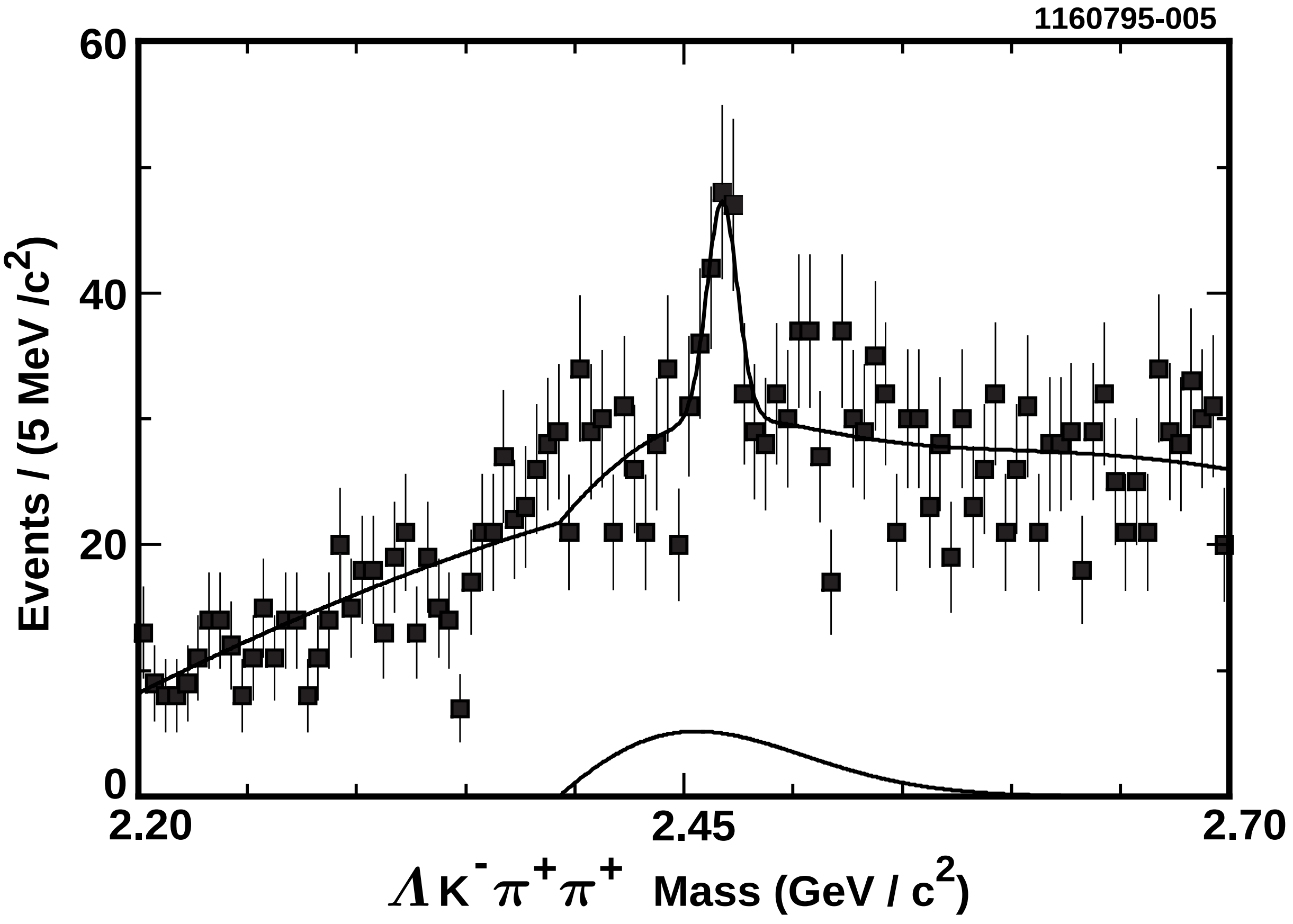

FIG. 4. Invariant mass distribution for $\Xi_c^+ \to \Lambda K^- \pi^+ \pi^+$. The curve at the bottom corresponds to the reflection peak from misidentified $\Lambda_c^+ \to \Lambda \pi^+ \pi^- \pi^+$ events.

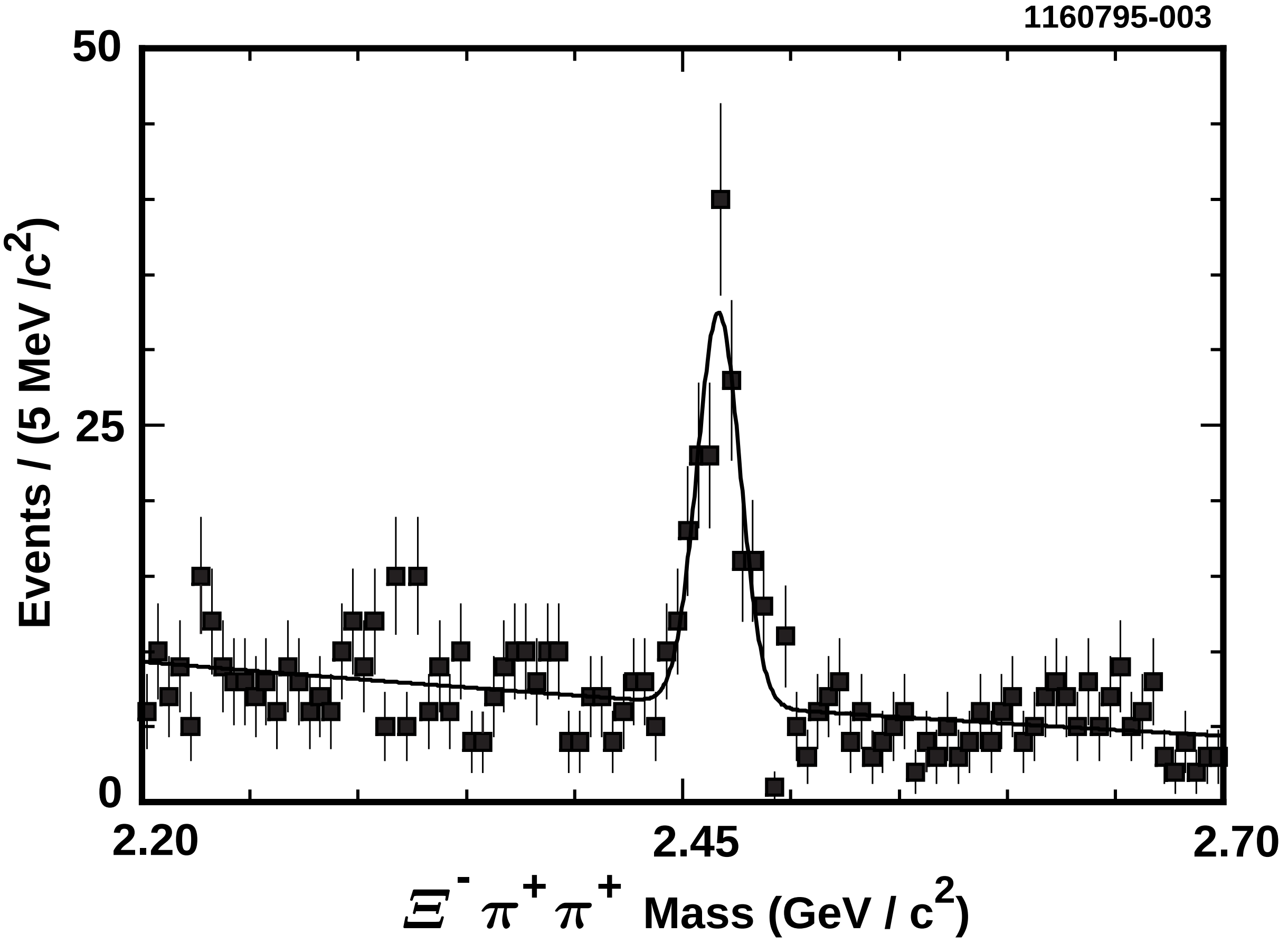

FIG. 5. Invariant mass distribution for $\Xi_c^+ \to \Xi^- \pi^+ \pi^+$.